\documentclass{icrc2009}

\usepackage{graphicx}   
\usepackage[caption=false]{caption}    
\usepackage[font=footnotesize]{subfig} 
\usepackage{fixltx2e}
\usepackage{url}

\newcommand{\shorttitle}[1]%
{\markboth{Proceedings of the 31\MakeLowercase{$^{st}$} ICRC, {\L}\'{o}d\'{z} 2009}{#1} }
\newcommand{\etal}{\MakeLowercase{\textit{et al. }}} 

\begin{document}
\title{The MAGIC Data Center}

\author{\IEEEauthorblockN{Ignasi Reichardt\IEEEauthorrefmark{1}\IEEEauthorrefmark{5},
    Javier Rico\IEEEauthorrefmark{1}\IEEEauthorrefmark{2}\IEEEauthorrefmark{5},
    Emiliano Carmona\IEEEauthorrefmark{3},
    Jose Luis Contreras\IEEEauthorrefmark{4},
    Juan Cortina\IEEEauthorrefmark{1},\\
    Roger Firpo\IEEEauthorrefmark{7},
    Llu\'{\i}s Font\IEEEauthorrefmark{6},
    Abelardo Moralejo\IEEEauthorrefmark{1},
    Daniel Nieto\IEEEauthorrefmark{4},
    Igor Oya\IEEEauthorrefmark{4},
    Raquel de los Reyes\IEEEauthorrefmark{4}\IEEEauthorrefmark{8},
    \\on behalf of the MAGIC collaboration}
  \\
\IEEEauthorblockA{\IEEEauthorrefmark{1}Institut de F\'{\i}sica d'Altes Energies, Edifici C7, Campus UAB, E-08193 Bellaterra, Spain}
\IEEEauthorblockA{\IEEEauthorrefmark{2}ICREA, E-08010 Barcelona, Spain}
\IEEEauthorblockA{\IEEEauthorrefmark{3}Max-Planck-Institut f\"ur Physik, D-80805 M\"unchen, Germany}
\IEEEauthorblockA{\IEEEauthorrefmark{4}Universidad Complutense, E-28040 Madrid, Spain}
\IEEEauthorblockA{\IEEEauthorrefmark{7}Port d'Informaci\'o Cient\'{\i}fica, Edifici D, Campus UAB, E-08193 Bellaterra, Spain}
\IEEEauthorblockA{\IEEEauthorrefmark{6}Universitat Aut\`onoma de Barcelona, E-08193 Bellaterra, Spain}
\IEEEauthorblockA{\IEEEauthorrefmark{8}presently at Max-Planck-Institut fur Kernphysik, P.O. Box 103980, D-69029 Heidelberg, Germany}
\IEEEauthorblockA{\IEEEauthorrefmark{5}Corresponding authors: ignasi@ifae.es, jrico@ifae.es }}

\shorttitle{I. Reichardt, J. Rico \etal The MAGIC Data Center}
\maketitle

\begin{abstract}
The MAGIC I telescope produces currently around 100TByte of raw data per year that is calibrated and reduced on-site at the Observatorio del Roque de los Muchachos (La Palma). Since February 2007 most of the data have been stored and further processed in the Port d'Informaci\'o Cient\'{\i}fica (PIC), Barcelona. This facility, which supports the GRID Tier 1 center for LHC in Spain, provides resources to give the entire MAGIC Collaboration access to the reduced telescope data. It is expected that the data volume will increase by a factor 3 after the start-up of the second telescope, MAGIC II. The project to improve the MAGIC Data Center to meet these requirements is presented. In addition, we discuss the production of high level data products that will allow a more flexible analysis and will contribute to the international network of astronomical data (European Virtual Observatory). For this purpose, we will have to develop a new software able to adapt the analysis process to different data taking conditions, such as different trigger configurations or mono/stereo telescope observations.
  \end{abstract}

\begin{IEEEkeywords}
massive data processing, GRID, European Virtual Observatory
\end{IEEEkeywords}
 
\section{Introduction}
MAGIC are two imaging atmospheric Cherenkov telescopes for gamma-ray astronomy located at the \textit{Observatorio del Roque de los Muchachos} (European Northern Observatory, La Palma island, Spain). The production of a large amount of data during normal operation is inherent to the observational technique. The storage and processing of these data is a technical challenge which the MAGIC collaboration has solved by profitting from infrastructures like those developed for LHC experiments.\\
During the last years, the MAGIC groups at IFAE and UAB (Barcelona) and UCM (Madrid) have set up, in collaboration with PIC, the MAGIC Data Center. The facility became operational in February 2007 and, as of now, is equipped with the needed storage resources and computing capabilities to process the data from the first telescope and make them available to the MAGIC Collaboration. However, as MAGIC II is expected to start generating data this year, we expect the data volume to be increased by a factor of 3 with respect to the present single telescope situation. In consequence, we will increase the capabilities of the Data Center by providing it with the needed hardware and human resources to make it able to centralise the storage and analysis of the data. The main goals of this extension are to allow fast massive (re-)processings of all stored data and to support data analysis for all MAGIC collaborators.\\
In what follows we will describe the present status of the Data Center and the foreseen upgrades required to deal with the data flow from the two-telescopes system. This will lead us to provide additional services useful for the collaboration and also for the astrophysical community.\\
 
\begin{table*}[th]
\begin{center}
\small
\begin{tabular}{lrrrr} 
\hline \hline
Telescope system		   &\bf MAGIC\,I&\bf MAGIC\,II &\bf MAGIC\,I+II\\
\hline
\# of pixels			   &577		&1081	      &1658     \\
\# of samples			   &50		&50	      &50     \\
Bytes per saple		   &2		&2	      &2      \\
Event size (kByte)		   &60.7	&110.0	      &170.7    \\
Event rate (Hz) 		   &350		&350	      &350      \\
RAW data volume (MByte/s)	   &20.8	&37.6	      &58.4     \\
RAW data volume (GByte/h)	   &73.0	&132.1	      &205.1    \\
Observation time per year (h)		   &1500	&1500	      &1500     \\
\bf RAW data volume (TByte/yr)	   &106.9	&193.6	      &300.5\\
Gzip compress factor on RAW data   &0.3		&0.3	      &0.3      \\
\bf Gzipped RAW data volume (TByte/yr) &32.1	&58.1	      &90.2\\
Calibration reduction factor      &0.034	&0.034        &0.034    \\
\bf CALIB data volume (TByte/yr)   &3.6 	&6.6	      &10.2 \\
Star reduction factor   	   &0.0047	&0.0026       &0.0033   \\
Star data volume (TByte/yr)	   &0.5		&0.5	      &1.0      \\
\bf REDUCED data volume (TByte/yr) & ---	&---	      &1.1  \\
\hline \hline
\end{tabular} 
\caption{Data volume for the different phases of the MAGIC telescope's analysis chain}
\label{tab:volume} 
\end{center}
\end{table*} 

\begin{table*}[!h]
  \begin{center}
\small
\begin{tabular}{lccc} 
\hline \hline
Phase  & Program & Input file & Output file \\
\hline
Data acquisition    &   DAQ                  & ---     & RAW \\
Calibration         & callisto               & RAW     & CALIB \\
Reduction           & star+superstar+melibea & CALIB   & REDUCED \\
High level analysis &  others                & REDUCED & others \\
\hline \hline
\end{tabular} 
\caption{The phases of MAGIC standard analysis, with their associated standard programs and input/output data types}
\label{tab:analysis} 
\end{center}
  \end{table*}
\normalsize

\section{MAGIC data: production and analysis}
The MAGIC telescopes have in their focal plane a camera segmented into a number \textit{c} of different pixels (each equipped with a photo-multiplier), whose signals are digitized by the DAQ system~\cite{signalreco}. Currently, MAGIC I is in service as a single telescope with a camera of 577 pixels. MAGIC II is under commissioning, and has an improved camera with 1039 pixels for regular operation plus 42 additional pixels equiped with experimental high quantum efficiency photodetectors for test purposes~\cite{hpd}\cite{camera}. \\
For every trigger, the single pixel signal is sampled \textit{s} times. Each sample is digitized with 12 bit precision and the resulting values stored in 2 byte fields. The information is then saved into a RAW data file. The size of a MAGIC RAW event is given by \textit{h} + 2byte\textit{$\cdot$s$\cdot$c}, where \textit{h} is a fixed-size (4.5 kByte) header describing the event. The event rate depends on the observation conditions and trigger configuration, and can range between 200 and 700 Hz. The average event rate during Observation Periods 67-73 (May-Dec 2008) was 350Hz. We will use this value to compute the data volume and storage needs summarized in Table~\ref{tab:volume}. \\
RAW event files are processed using the MAGIC standard Analysis and Reconstruction Software (MARS)~\cite{Abe}. The first step is a program dubbed \textit{callisto}, which calibrates the Cherenkov pulse's intensity and arrival time, producing a so-called CALIB data file. This part of the analysis is the most CPU demanding, therefore CALIB data files are saved before further data processing. The rest of the analysis chain consists of a set of executables taking as input the output of the previous program in the chain: \textit{star} computes the parameters describing the Cherenkov images of the individual telescope~\cite{hillas}\cite{timinganl}; \textit{superstar} merges the information of a given shower from the two telescopes; and \textit{melibea} computes the estimated energy, arrival direction and the so-called \textit{hadronness} (a parameter used for gamma/hadron discrimination~\cite{rf}). The output from \textit{star} and \textit{melibea} will be referred to as different steps REDUCED data files, and is also stored permanently.\\
Estimations of the data volume at the different stages of the analysis chain, and for the different telescope configurations are also shown in Table~\ref{tab:volume}, The different phases of the standard analysis, the name of the standard programs and the input and output file formats are summarized in Table~\ref{tab:analysis}.\\
A diferent route is followed by the Monte-Carlo simulated events that are used for the estimation of the energy and \textit{hadronness}. In this case, instead of RAW files, atmospheric particle showers are generated using CORSIKA and then digested in two steps (\textit{reflector} and \textit{camera}) that finally produce data-like files that are calibrated and reduced with the same programs used for real data. The parameters of the detector simulation are adapted to the telescope performance in different observation periods and configurations. Therefore, several versions of the Monte-Carlo library are provided at the Data Center. Presently, the simulated events are generated at the INFN Padova and Udine, and the resulting files require 10TByte of disk space.\\

\section{Description of the Data Center services}
\label{sec:needs}
 Currently, the MAGIC Data Center takes care of the following tasks:
\begin{itemize}
\item Data transfer from La Palma to PIC, via internet and tapes
\item Data storage on tapes and disk at PIC
\item Data access at all data processing levels (RAW, CALIB, REDUCED) for all MAGIC collaborators
\item Real-time, automatic analysis of the data, processed with MAGIC standard software
\item Reanalysis of all stored data in case of software updates and bugfixes
\item MAGIC data base
\item Software repository and bug tracker
\item Storage of the data quality control files
  \end{itemize}
During normal data taking on the site, RAW data are stored into disk, and later recorded to tape. The tapes are then sent from La Palma to PIC via airmail, since currently there is not enough Internet bandwidth between the island and Europe to support the transfer of the files.\\
The computing system on the site also performs the so-called \textit{OnSite}~\cite{Igor} analysis, by which RAW data are processed, producing CALIB files and REDUCED files right after the data taking. These files are indeed transferred to PIC via Internet, together with some log files generated by the subsystems of the telescope.\\
Currently, all tapes received at PIC are downloaded to a buffer disk and then written back to tape grouped by source and observation night in a single file (an \textit{ISO volume}) that can be mounted as an external unit in the file system. This procedure is obsolete and recently has been found more optimal to store the files directly in tapes.\\
The data are organized in a data base and served to the collaboration through a user interface machine and a web site. Until late 2008 these data were in a NFS file system, but currently, all data are being migrated to a new GRID-based file system (dCACHE) that allows more transparent access (in the sense of making no difference between tape and disk storage) to any level of RAW and processed data. It is planned that in July 2009 the NFS will be finally dismantled and all applications and data access will be based on GRID.\\
The reduction of the files that arrive via Internet is triggered by scripts that run automatically every few minutes and notice when transfers have successfully finished. When this happens, batch jobs are generated and submitted to the GRID with a specific configuration that ensures that they will run at PIC (the only Computing Element where MARS is currently installed). This set of scripts can also be used to massively reprocess all the data stored at PIC in case that a bug is found in the software or some improvement in the analysis makes it worth. Massive operations of this kind have been performed twice in 2008 and once more in 2009. For the last one, we have estimated that we used 115 days of CPU time in two weeks. This means that we have the capability to run \textit{star} on one year of data in about 8 days. It is worth to mention that this peak processing rate would have never been achieved with just the minimum number of CPU cores that the MAGIC project is granted at PIC according to its share. In fact, we should always have access to at least 7 cores any time, but in periods of low usage from other experiments we have got up to 10 times these resources.\\
Finally the Data Center also provides a ``Concurrent Version Server'' (\textit{CVS}) for the software development and the Daily Check, which generates a daily report on the data quality and the telescope stability. \\

\section{Future plans for the Data Center}
In a near future we want to provide additional services to allow a more agile analysis by any MAGIC collaborator and also easier access of anyone to published data. For this we intend to:
\begin{itemize}
\item Extend the automatic data reduction up to \textit{melibea}
\item Provide resources and tools for high level analyses by any MAGIC collaborator
\item Open the MAGIC public data to the whole scientific community by linking it to the European Virtual Observatory
\end{itemize}
Currently, the high level analysis, starting from Melibea, is carried out by analysers that select a MC-gamma sample and a real data sample fitting well the observational conditions of the analyzed data. These samples are used in the training of the multidimensional technique of selection of gamma-like events (the Random Forest method~\cite{rf}). We intend to automatize also this part of the analysis at PIC in the near future, making the task of the analyzer simpler.\\
Also the computing power of PIC can be more widely exploited by opening the job submission to the rest of the collaboration. The already working roles of the GRID scheme allow to assign priorities to the CPU farm users according to their duties, securing that the official data reduction is not delayed.\\
Finally, we intend to establish a link with the European Virtual Observatory in order to share potentially interesting data for the astrophysical community. For this purpose a software that will translate ROOT information in the widely used format in astronomy - FITS is currently being developed.\\

\section{Conclusion}
The MAGIC Data Center based at PIC is already providing quality services to the MAGIC collaboration, exploiting when possible the extra resources that a GRID-based infrastructure implies.\\
The success of the two year experience as official Data Center makes us push for the extension of the current facilities. We hope this will improve the access to the data by the MAGIC analyzers, and will make it more transparent for interested astrophysicists outside the collaboration.\\

\end{document}